\documentclass[12pt]{article}
\usepackage[dvips]{color}
\usepackage{graphicx}
\usepackage{amsthm,amssymb,amsmath,graphics,times}

\usepackage[colorlinks=true, linkcolor=blue, citecolor=blue,urlcolor=blue, breaklinks=true]{hyperref}

\numberwithin{equation}{section}
\newcommand{\RR}{\rm I\kern -1.6pt{\rm R}}

\newtheorem{thm}{Theorem}[section]

\date{}

\def\mv{\mathcal{V}}
\def\mr{\mathcal{R}}

\def\Om{\Omega}
\def\Ld{\Lambda}

\def\ds{\displaystyle}

\def\bc{\begin{center}}\def\ec{\end{center}}
\def\ba{\begin{aligned}}\def\ea{\end{aligned}}

\begin{document}



\centerline{\Large \sffamily Theory of Infectious Diseases with Testing}

\centerline{\Large \sffamily and Testing-less Covid-19 Endemic}

\medskip
\centerline{\sffamily  Bo Deng\footnote{Department of Mathematics,
University of Nebraska-Lincoln, Lincoln, NE 68588. Email: {\tt
bdeng@math.unl.edu}} and Chayu Yang \footnote{Department of Mathematics,
University of Nebraska-Lincoln, Lincoln, NE 68588. Email: {\tt
chayuyang@unl.edu}} }

\noindent
\textbf{Abstract: What is the long term dynamics of the Covid-19 pandemic? How will it end? Here we
constructed an infectious disease model with testing and analyzed the existence and stability
of its endemic states. For a large parameter set, including those relevant to the SARS-CoV-2
virus, we demonstrated the existence of one endemic equilibrium without testing and one endemic
equilibrium with testing and proved their local and global stabilities for some cases.
Our results suggest that the pandemic is to end with a testing-less endemic
state through a novel and surprising mechanism called stochastic trapping.}

\bigskip


Testing for infectious diseases for this paper is defined to be any diagnosis which
results in case numbers on record. A test-positive patient may become recovered or
dead later. Without testing the epidemiological dynamics of an outbreak is a known-unknown.
It is through testing that we gain a window to the spread of the disease in speed and in
scope. Because of such an essential role testing plays in epidemiological understanding, 
theoretical models should consider testing as an important compartment.

In this paper, we first introduce a model modified from the SICM model
with testing from \cite{Deng23}. For the purpose to understand long term behaviors
of infectious diseases, we incorporate both the natural birth rate and the natural death
rate into the new model, which is referred to as the SICMR model for distinction. We then
obtain the existence as well as local and global stabilities of endemic equilibrium states,
which are of two types: endemic equilibrium without testing and endemic equilibrium
with testing. As an application, we apply our theory to the U.S. Covid-19 pandemic by
first best-fitting the model to the case and death numbers, and then analyzing
the long term behaviors of the best-fitted model. To our surprise, we find that
our model is capable of deducing that the endemic state without large scale testing
is the outcome. This happens either because the endemic state has insignificant numbers of testing or due to the fact that the model has a testing-free
invariant manifold and the SARS-CoV-2's outbreak trajectories tend to fall towards
an exponentially attracting region of the manifold, and as a result, when a
trajectory stays long enough near the trapping region, stochastic fluctuations
will eventually push the trajectory into the testing-free zone for good.

\section{SICMR Model}

All epidemiological modeling starts with the basic SIR model (\cite{Brau2013,Driessche})
in a fixed population
\[
S'=-cSI,\ I'=cSI-r I,\  R'=r I, \  S+I+R=N
\]
where $S$ is the number of susceptible at time $t$, $I$ the infected,
$R$ the recovered, and $S'=S'(t)$ is the rate of change
(derivative) of variable $S(t)$. Parameter $c$ is the per-infected infection rate
and $r$ is the recovery rate, $N$ is the fixed population. Two modifications
are proposed in \cite{Deng23}. The first is \textit{the openness hypothesis} that
$N$ is not fixed at the total population of a geographic region or state,
say the U.S., but instead, it is a parameter, referred to as \textit{the effective susceptible population} for a period of time. This is because, for example, when SARS-CoV-2 first appeared in Seattle, WA, it did not make every people in Nebraska susceptible. Also, if a person isolates themselves in terms of mitigation
from the population, they can't be susceptible to the disease at such times.
The second modification is the inclusion of testing because
it is the case number and death number from the disease that we can see not
the $S$ or $I$, which are not directly observable and can only be
triangulated by the testing numbers. The inclusion of testing results
two more compartments: the class $C$ for confirmed by testing, and the
class $M$ for monitored after confirmation that requires at least
one more testing before going into the recovered class. The new model
for this paper includes three more elements: the natural birth and death
rates, and the intrinsic recovery rate. The dimensionless version of the
model is as follows
\begin{equation}\label{SICMR}
\begin{cases}
\;S'&=\alpha-cSI-\mu S\\
\;I'&=cSI-\frac{pC}{\varepsilon+I+aM}I-\gamma I-\mu I\\
\;C'&=\frac{pC}{\varepsilon+I+aM}I-mC-dC-\mu C\\
\;M'&=mC-qM-\mu M\\
\;R'&=qM+\gamma I-\mu R\\
\end{cases}
\end{equation}
Here the unit for time is in day, $\alpha$ is the influx rate,
approximately the natural per-capita daily birth and
immigration rate, $c$ is the daily infection rate per
infectious person, $\mu$ is the efflux rate, i.e. the
natural per-capita daily
death and emigration rate,  $\gamma$ is the product
of the recovered rate, $r$, and the proportion of
those infected but not tested. $M$ is the class of
test-positive individuals who will eventually recover from
the infection and who will receive at least another test before
being put into the recovered category at the monitored recovery
rate $q$. This class of individuals is taken out from the
infective class $I$ by themselves or institutionalized
isolation. Parameter $m$ is the monitoring rate with
which test-positive individuals are put into the
monitored class $M$. Parameter $d$ is the death rate
of those who are tested positive and eventually die from the disease.

Parameters $p,\varepsilon,a$ are all related
to the daily test-positive rate $pCI/(\varepsilon+I+aM)$. It
is the Holling's Type II functional form (\cite{Holl1959,Deng23})
from theoretical ecology. Holling's theory, derived for
predation, is universal to all processes involving two entities
one of which must take time to change the encountering of both
into something else. In our setting, disease testing is an agent
or infrastructure which is to find out infected individuals by diagnostic
interaction before putting them into the confirmed class $C$. Testing
is also the means to find out if an infected individual
under monitoring is no longer infectious and thus can be released to the
recovered class $R$. For the first class, there is a discovery
probability rate $a_1$ of the infected class $I$ that will be tested
and confirmed. For the second class, there is a repeating test
rate $a_2$ which is the average number of tests an individual
will receive over an average period of days under monitoring.
For both cases, there is an average time $h$ needed to complete a test.
Under these assumptions, the number of daily cases confirmed is
the following Holling Type II function
\[
\hbox{$\frac{a_1\bar I}{1+a_1 h \bar I+a_2 h \bar M}\bar C =
\frac{(1/h) \bar C/N_0}{1/(a_1hN_0) + \bar I/N_0+(a_2/a_1) \bar M/N_0}\bar I
=\frac{pC}{\varepsilon+I+aM}\bar I$}
\]
where $p=1/h$ is the rate of testing and $h$ is testing time,
$a=a_2/a_1$ is the ratio of testing rates for monitored and infected,
and $\varepsilon=1/(a_1hN_0)$ with $N_0$ being the effective susceptible
population. $\bar I,\bar C,\bar M$ are dimensional variables and
$I=\bar I/N_0, C=\bar C/N_0$, etc. are dimensionless variables.
Because $a_1h$ is moderate and $N_0$ is large, we will keep $\varepsilon$ to be a small parameter. Alternatively, one can start
with the assumption that the daily confirmed number is proportional to
the product of the infected and the confirmed because one class
has positive feedback on the other class, the so-called Holling Type I
functional form. Because testing takes time,
therefore, the daily rate must be constrained by the time allowed
and the constraining factor is exactly in the form of the denominator
by Holling's theory. See \cite{Deng23} for more explanations on
the functional form.

To convert the dimensionless model (\ref{SICMR}) into its dimensional one, just multiply the equation by the parameter $N_0$ and replace $\bar S=S\times N_0,
\bar I=I\times N_0$ etc., and all the parameters remain the
same except for that $\alpha$ is replaced by $\alpha N_0$ and $c$ is
replaced by $c/N_0$. For the dimensionless model, we assume the
initial values sum up approximately equal to 1: $S(0)+I(0)+C(0)+M(0)+R(0)\simeq 1$.
If we rewrite the daily testing rate as $P=\frac{C}{\varepsilon+I+aM}pI$,
then the factor $\frac{C}{\varepsilon+I+aM}$ is the ratio that infected are
tested and the complement $1- \frac{C}{\varepsilon+I+aM}$ is the fraction
of the infected class going directly into the recovered class $R$ with
the natural recover rate $r$. To keep the model simple, we will use a parameter,
namely $\gamma$, for the product. Last, we note that the SICM model of
\cite{Deng23} is the system (\ref{SICMR}) without all the terms with
the Greek letter parameters, $\alpha,\mu,\varepsilon,\gamma$. Also note
that the equation $R$ is decoupled from the rest, which can make
analysis and computation easier.

By adding the five equations in \eqref{SICMR}, we have
\[
(S+I+C+M+R)'\le \alpha-\mu(S+I+C+M+R).
\]
Hence, we get an invariant set below for model \eqref{SICMR}
\[
\Om=\{(S, I, C, M, R)\in\mathbb{R}_+^5 : S+I+C+M+R\le\frac{\alpha}{\mu}\}.
\]
If $I=0$, we can obtain a unique disease-free equilibrium
\[
E_0=(\frac{\alpha}{\mu}, 0, 0, 0, 0, 0).
\]
In this model, we consider $I, C$, and $M$ to be
all infectious compartments. Using the next-generation matrix
method \cite{Driessche}, we choose $cSI, \frac{pC}{\varepsilon+I+aM}I$, and $mC$ to be the rates of appearance of new infections and $\frac{pC}{\varepsilon+I+aM}I, (m+d+\mu) C,$ and $(q+\mu) M$ to be the rate of transfer of individuals out of each infectious compartment $I, C$, and $M$, respectively. Then, taking the partial derivatives of those rates with respect to variables $I, C$, and $M$, we obtain the new infection matrix $F$ and the transitive matrix $V$ at the disease-free equilibrium $E_0$:
\[
F=\begin{pmatrix}\frac{c\alpha}{\mu}  & 0 & 0\\ 0 & 0 & 0 \\
0 & m & 0\end{pmatrix} ~\text{and}~ V=\begin{pmatrix}\mu+\gamma & 0 & 0\\
0 & m+d+\mu & 0 \\0 & 0 & q+\mu\end{pmatrix}.
\]
Hence, the next-generation matrix is given by
\[
FV^{-1}=\begin{pmatrix}\frac{c\alpha}{\mu(\mu+\gamma)} & 0 & 0\\
0 & 0 & 0\\0 & \frac{m}{m+d+\mu} & 0\end{pmatrix}.
\]
The basic reproduction number is defined by the spectral
radius of $FV^{-1}$, that is,
\begin{equation}\label{defR0}
\mathcal{R}_0=\rho(FV^{-1})=\dfrac{c\alpha}{\mu(\mu+\gamma)}.
\end{equation}
The endemic equilibrium $(S, I, C, M, R), I>0$ satisfies
\begin{align}
\alpha-cSI-\mu S&=0\label{ee1}\\
cS-\frac{pC}{\varepsilon+I+aM}-\gamma-\mu&=0 \label{ee2}\\
\frac{pC}{\varepsilon+I+aM}I-mC-dC-\mu C&=0 \label{ee3} \\
mC-qM-\mu M&=0\label{ee4}\\
qM+\gamma I-\mu R&=0 \label{ee5}
\end{align}
Equations \eqref{ee4} and \eqref{ee5} give $C=\frac{q+\mu}mM$
and $R=\frac1{\mu}(qM+\gamma I)$. By substituting $C=\frac{q+\mu}mM$
into equations \eqref{ee2} and \eqref{ee3}, we have
\begin{align}
cS-\frac{p(q+\mu)M}{m(\varepsilon+I+aM)}-\mu-\gamma&=0, \label{e2}\\
M\left(\frac{pI}{\varepsilon+I+aM}-m-d-\mu\right)&=0. \label{e3}
\end{align}
For $M=0$, system \eqref{SICMR} has a boundary equilibrium below if $\mr_0>1$.
\[
\ds E_1=(S_1, I_1, 0, 0, R_1)=\left(\frac{\mu+\gamma}c,
\frac{\mu}c(\mr_0-1), 0, 0, \frac{\gamma}{c}(\mr_0-1)\right).
\]
For $M\neq0$, by solving equations \eqref{ee1}-\eqref{e3}, we can find
\[
S_*=\frac{\alpha}{cb(M_*+\frac{\varepsilon}{a})+\mu},
I_*=b(M_*+\frac{\varepsilon}{a}), C_*=\frac{q+\mu}mM_*, R_*=\frac1{\mu}(qM_*+\gamma I_*),
\]
where $b=\frac{a(m+d+\mu)}{p-m-d-\mu}$ and $M_*$ is the positive root of the equation
\begin{equation}\label{eqM}
a_2M^2+a_1M+a_0=0,
\end{equation}
from \eqref{ee3} where
\[
\ba a_2&=\frac{(q+\mu)(m+d+\mu)}{m}+(\mu+\gamma)b,\\
 a_1&=\frac{(q+\mu)(m+d+\mu)}{m}(\frac{\varepsilon}a
 +\frac{\mu}{bc})+\frac{2\varepsilon b(\mu+\gamma)}{a}+\frac{\mu(\mu+\gamma)}{c}-\alpha,\\
a_0&=\frac{\varepsilon}a(\frac{\varepsilon b(\mu+\gamma)}{a}+\frac{\mu(\mu+\gamma)}{c}-\alpha)
=\frac{\varepsilon\mu(\mu+\gamma)}{ac}(\frac{\varepsilon bc}{a\mu}+1-\mr_0).
\ea
\]
Note that $a_1>\frac{a}{\varepsilon}a_0$. Hence, if $a_0\ge0$, that is,
$\mr_0\le1+\frac{\varepsilon bc}{a\mu}$, then there is no positive root for
equation \eqref{eqM} because $f(0)=a_0>0,f'(0)=a_1>a_0>0$ for
the quadratic polynomial $f(x)=a_2x^2+a_1x+a_0$. If $a_0<0$,
that is, $\mr_0>1+\frac{\varepsilon bc}{a\mu}$,
then there is a unique positive root $M_*$ for equation \eqref{eqM}. Thus,
if $p>m+d+\mu$ and $\mr_0> 1+\frac{\varepsilon bc}{a\mu}$, we have
$S_*, I_*, C_*, M_*, R_*$ are all positive, and hence, system \eqref{SICMR}
admits an interior equilibrium
\[
E_*=(S_*, I_*, C_*, M_*, R_*).
\]
In addition, when $E_1$ and $E_*$ both exist, we can show that the number
of infected individuals $I_*$ at the endemic equilibrium $E_*$ is less than
the number of infected individuals $I_1$ at the boundary equilibrium $E_1$.
In fact, let $\Delta=\mr_0-1-\frac{\varepsilon bc}{a\mu}>0$. Note that
\[
\ds I_*<I_1\Longleftrightarrow b(M_*+\frac{\varepsilon}a)<
\frac{\mu}c(\mr_0-1)\Longleftrightarrow M_*<\frac{\mu}{bc}(\mr_0-1)
-\frac{\varepsilon}a\Longleftrightarrow M_*<\frac{\mu}{bc}\Delta.
\]
Since $a_2M_*^2+a_1M_*+a_0=0$, it suffices to show
that $a_2(\frac{\mu}{bc}\Delta)^2+a_1(\frac{\mu}{bc}\Delta)+a_0>0$.
By direct algebra calculation, one can verify that
\[
\ba &a_2(\ds\frac{\mu}{bc}\Delta)^2+a_1(\frac{\mu}{bc}\Delta)+a_0\\
=&\left(\frac{(q+\mu)(m+d+\mu)}m+(\mu+\gamma)b\right)\left(\frac{\mu}{bc}\Delta\right)^2
+\left(\frac{(q+\mu)(m+d+\mu)\mu}{mbc}\left(1+\frac{\varepsilon bc}{a\mu}\right)\right.\\
&+\left.\frac{\varepsilon b(\mu+\gamma)}a-\frac{\mu(\mu+\gamma)}{c}\Delta\right)\frac{\mu}{bc}\Delta
-\frac{\varepsilon\mu(\mu+\gamma)}{ac}\Delta\\
=&\left[\left(\frac{(q+\mu)(m+d+\mu)}m+(\mu+\gamma)b\right)\frac{\mu}{bc}\Delta
+\frac{(q+\mu)(m+d+\mu)\mu}{mbc}(\mr_0-\Delta)\right.\\
&-\left.\frac{\mu(\mu+\gamma)}{c}\Delta\right]\frac{\mu}{bc}\Delta\\
=&\left[\left((\frac{(q+\mu)(m+d+\mu)}m+(\mu+\gamma)b)\frac{\mu}{bc}
-\frac{(q+\mu)(m+d+\mu)\mu}{mbc}-\frac{\mu(\mu+\gamma)}{c}\right)\right.\Delta\\
&+\left.\frac{(q+\mu)(m+d+\mu)\mu}{mbc}\mr_0\right]\frac{\mu}{bc}\Delta\\
=&\frac{(q+\mu)(m+d+\mu)\mu^2}{m(bc)^2}\mr_0\Delta\\
>&0.
\ea
\]
We summarize the above results in the following theorem.
\begin{thm}\label{ThmExist}
For system \eqref{SICMR},
\begin{itemize}
\item [1.] there always exists a unique disease-free equilibrium $E_0$;
\item [2.] if $\mr_0>1$, a unique boundary equilibrium $E_1$ occurs;
\item [3.] if $p\le m+d+\mu$, there is no interior equilibrium;
\item [4.] if $p>m+d+\mu$ and $\mr_0>1+\frac{\varepsilon bc}{a\mu}$,
a unique interior equilibrium $E_*$ exists. Furthermore, $I_*<I_1$.
\end{itemize}
\end{thm}

For stability of the endemic equilibria we have the following result
\begin{thm}\label{ThmStab}
For system \eqref{SICMR},
\begin{itemize}
\item [1.] the disease-free equilibrium $E_0$ is globally asymptotically stable in $\Om$ for
$\mr_0\le1$ and it is unstable for $\mr_0>1$;
\item [2.] if $p<m+d+\mu$ and $1<\mr_0<1+\frac{\varepsilon bc}{a\mu}$, the boundary equilibrium $E_1$ is unstable;
\item [3.] if $p<m+d+\mu$ and $\mr_0\ge1+\frac{\varepsilon bc}{a\mu}$, the boundary equilibrium $E_1$ is
globally asymptotically stable in $\Om/\{S\hbox{-axis}\}$;
\item [4.] if $p>m+d+\mu$ and $1<\mr_0\le1+\frac{\varepsilon bc}{a\mu}$, the boundary equilibrium
$E_1$ is globally asymptotically stable in $\Om/\{S\hbox{-axis}\}$
\item [5.] if $p>m+d+\mu$ and $\mr_0>1+\frac{\varepsilon bc}{a\mu}$, the boundary equilibrium $E_1$
is unstable.
\end{itemize}
\end{thm}
\begin{proof}
The Jacobian of system \eqref{SICMR} is
\[
J=\begin{pmatrix}-cI-\mu & -cS & 0 & 0 &0\\
cI & cS-\frac{pC(aM+\varepsilon)}{(\varepsilon+I+aM)^2}-\gamma-\mu & -\frac{pI}{\varepsilon+I+aM}
 & \frac{apCI}{(\varepsilon+I+aM)^2} & 0 \\
0 & \frac{pC(aM+\varepsilon)}{(\varepsilon+I+aM)^2} & \frac{pI}{\varepsilon+I+aM}-m-d-\mu &
-\frac{apCI}{(\varepsilon+I+aM)^2} & 0 \\
0 & 0 & m &-q-\mu & 0\\
0 & \gamma & 0 & q & -\mu
\end{pmatrix}.
\]

1. For $\mr_0>1$, it follows from the Jacobian at $E_0$
\[
J(E_0)=\begin{pmatrix}-\mu & -\frac{c\alpha}{\mu} & 0 & 0 &0\\
0 & (\mu+\gamma)(\mr_0-1) & 0 & 0 & 0 \\ 0 & 0 & -m-d-\mu & 0 & 0 \\
0 & 0 & m &-q-\mu & 0\\
0 & \gamma & 0 & q & -\mu  \end{pmatrix}
\]
has a positive eigenvalue $(\mu+\gamma)(\mr_0-1)$ that $E_0$ is unstable.

For $\mr_0\le1$, consider a Lyapunov function
\[
\ds\mv=I+C+M.
\]
In $\Om$, it is easy to see that
\[
\ba\mv'&=cSI-(\gamma+\mu)I-(d+\mu)C-(q+\mu)M\\
&\le(\frac{c\alpha}{\mu}-\gamma-\mu)I-(d+\mu)C-(q+\mu)M\\
&\le(\mu+\gamma)(\mr_0-1)I-(d+\mu)C-(q+\mu)M\\
&\le0
\ea
\]
Note that $\mv'=0$ implies $S=\frac{\alpha}{\mu}$ and $C=M=0$, and the
largest positive invariant subset of the set $\{(S, I, C, M, R)\in\Om:
S=\frac{\alpha}{\mu}, C=M=0\}$ is  the disease-free equilibrium $\{E_0\}$.
Thus, $\{E_0\}$ is the largest positive invariant set on
$\{(S, I, C, M, R)\in\Om : \mv'=0\}$. By LaSalle invariant
principle \cite{LaSalle}, $E_0$ is globally asymptotically stable in $\Om$.

2. The Jacobian of system \eqref{SICMR} at $E_1$ is
\[
J(E_1)=\begin{pmatrix}-cI_1-\mu & -cS_1 & 0 & 0 &0\\ cI_1 & 0 &
-\frac{pI_1}{\varepsilon+I_1} & 0 & 0 \\ 0 & 0 & \frac{pI_1}{\varepsilon+I_1}-m-d-\mu
& 0 & 0 \\ 0 & 0 & m &-q-\mu & 0\\ 0 & \gamma & 0 & q & -\mu  \end{pmatrix}
\]
where the eigenvalue $\frac{pI_1}{\varepsilon+I_1}-m-d-\mu
=\frac{\mu(p-m-d-\mu)}{c(\varepsilon+I_1)}(\mr_0-1-\frac{\varepsilon bc}{a\mu})$.
Hence, for $p<m+d+\mu$, it is positive for $1<\mr_0<1
+\frac{\varepsilon bc}{a\mu}$ and $E_1$ is unstable.

3. Similar to the global stability analysis in \cite{YLMW}, we note that
$ S'\le \alpha-cSI-\mu S, \ I'\le cSI-\gamma I-\mu I$ for system (\ref{SICMR})
and the following SIR system
\[
\begin{cases}
S'=\alpha-cSI-\mu S,\\
I'=cSI-\gamma I-\mu I
\end{cases}
\]
has a global attractor $(S_1, I_1)$
in $\mathring{\Om}_1$ for $\mr_0>1$, where $\Om_1=\{(S, I)
=\mathbb{R}_+^2 : S+I\le\frac{\alpha}{\mu}\}$.
By comparison theorem \cite{Smith} ,
we have $\ds\limsup S\le S_1$ and $\ds\limsup I\le I_1$. Hence, we may
consider the following attracting set in $\Om$ for model (\ref{SICMR})
\[
\Gamma=\{(S, I, C, M, R)\in\mathbb{R}_+^5 : S+I+C+M+R\le\frac{\alpha}{\mu}, S\le S_1, I\le I_1\}.
\]
Let the Lyapunov function
\[
\mv_1=C.
\]
Then we have
\[
\ba\mv'_1=&\frac{pCI}{\varepsilon+I+aM}-mC-dC-\mu C\\
\le&\frac{pCI_1}{\varepsilon+I_1}-(m+d+\mu)C\\
=&\frac{\mu(p-m-d-\mu)}{c(\varepsilon+I_1)}(\mr_0-1-\frac{\varepsilon bc}{a\mu})C\\
\le&0
\ea
\]
for $\mr_0\ge1+\frac{\varepsilon bc}{a\mu}$. Since $\mv'_1=0$ implies
either $I=C=0$ or $I=I_1$ and $M=0$, the largest positive invariant
subset of the set $\{(S, I, C, M, R)\in\Gamma: \mv'_1=0\}$ is $\{E_0, E_1\}$.
By LaSalle invariant principle, all solution curves in $\Gamma$ will
approach either $E_0$ or $E_1$. Note that $E_0$ only attracts the solution
curve on the $S$\hbox{-axis} for $\mr_0>1$. Hence, all other solution curves
in $\Gamma/\{S\hbox{-axis}\}$ will approach $E_1$, which proves
that $E_1$ is globally asymptotically stable in $\Om/\{S\hbox{-axis}\}$.

4. Similarly, by using the same Lyapunov function $\mv_1=C$, one can obtain that
$E_1$ is globally asymptotically stable in $\Om/\{S\hbox{-axis}\}$
for $1<\mr_0\le1+\frac{\varepsilon bc}{a\mu}$.

5. If $p>m+d+\mu$, then $E_1$ is unstable for
$\mr_0>1+\frac{\varepsilon bc}{a\mu}$ since the eigenvalue
$\frac{pI_1}{\varepsilon+I_1}-m-d-\mu$ is positive.
\end{proof}

\section{Simplified SICMR Model}

For comparison purposes and to understand the global stability of the interior endemic equilibrium
$E_*$ better, we consider a simplified model by using Holling's Type I form for the testing rate:
\begin{equation}\label{s-sicm}
\begin{cases}
\; S'&=\alpha-cSI-\mu S,\\
\; I'&=cSI-pCI-\gamma I-\mu I,\\
\; C'&=pCI-mC-dC-\mu C,\\
\; M'&=mC-qM-\mu M,\\
\; R'&=qM+\gamma I-\mu R.
\end{cases}
\end{equation}
We consider model \eqref{s-sicm} in the invariant set
\[
\Om=\{(S, I, C, M, R)\in\mathbb{R}_+^5 : S+I+C+M+R\le\frac{\alpha}{\mu}\}.
\]
Model \eqref{s-sicm} exists a unique disease-free equilibrium
$E_0=(\frac{\alpha}{\mu}, 0, 0, 0, 0, 0)$ and the basic reproduction number is still
\[
\mathcal{R}_0=\dfrac{c\alpha}{\mu(\mu+\gamma)}.
\]
In addition, we can obtain a boundary equilibrium
\[
E_1=(S_1, I_1, 0, 0, R_1)=\ds\left(\frac{\mu+\gamma}c, \frac{\mu}c(\mr_0-1), 0, 0, \frac{\gamma}c(\mr_0-1)\right)
\]
for $\mr_0>1$ and an interior equilibrium
\[
E_*=(S_*, I_*, C_*, M_*, R_*)
\]
for $\mr_0>1+\frac{c(m+d+\mu)}{\mu p}$, where
\[
\hbox{$I_*=\frac{m+d+\mu}p, S_*=\frac{\alpha}{cI_*+\mu},
C_*=\frac{\mu(\mu+\gamma)}{p(cI_*+\mu)}(\mr_0-1-\frac{c(m+d+\mu)}{\mu p}),
M_*=\frac{mC_*}{q+\mu}$,  and $R_*=\frac{qM_*+\gamma I_*}{\mu}$.}
\]
Clearly, $\mr_0>1+\frac{c(m+d+\mu)}{\mu p}$ implies that $I_*<I_1$. Thus, we have
the following theorem.
\begin{thm}\label{ThmExist2} For system \eqref{s-sicm},
\begin{itemize}
\item [1.] there always exists a unique disease-free equilibrium $E_0$;
\item [2.] there exists a unique boundary equilibrium $E_1$ for $\mr_0>1$;
\item [3.] there is a unique interior equilibrium $E_*$ for $\mr_0>1+\frac{c(m+d+\mu)}{\mu p}$.
Furthermore, $I_*<I_1$.
\end{itemize}
\end{thm}

Similarly, for stability we have
\begin{thm}\label{ThmStab2}
For system \eqref{s-sicm},
\begin{itemize}
\item [1.] the disease-free equilibrium $E_0$ is globally asymptotically stable
in $\Om$ for $\mr_0\le1$ and it is unstable for $\mr_0>1$;
\item [2.] the boundary equilibrium $E_1$ is globally asymptotically stable
in $\Om/\{S\hbox{-axis}\}$ for $1<\mr_0\le1+\frac{c(m+d+\mu)}{\mu p}$,
and becomes unstable for $\mr_0>1+\frac{c(m+d+\mu)}{\mu p}$;
\item [3.] the interior equilibrium $E_*$ is globally asymptotically
stable in $\mathring{\Om}$ for $\mr_0>1+\frac{c(m+d+\mu)}{\mu p}$.
\end{itemize}
\end{thm}
\begin{proof}
By using the same proof in Theorem \ref{ThmStab}, it is easy to obtain
the stabilities of $E_0$ and $E_1$. We only prove (3) by
using the following Lyapunov function see (\cite{SD, YW, YWGW} in $\mathring{\Om}$
\[
\mathcal{V}_2=\ds\frac12(S-S_*)^2+S_*(I-I_*-I_*\ln\frac{I}{I_*}+C-C_*-C_*\ln\frac{C}{C_*}).
\]
It follows from $\alpha=cS_*I_*+\mu S_*, \gamma+\mu=cS_*-pC_*$, and $m+d+\mu=pI_*$ that
\[
\ba\mathcal{V}'_2=&(S-S_*)S'+S_*(\frac{I-I_*}II'+\frac{C-C_*}CC')\\
=&(S-S_*)(c(S_*I_*-SI)+\mu(S_*-S))+S_*\big((I-I_*)(c(S-S_*)-p(C-C_*))\\
&+(C-C_*)p(I-I_*)\big)\\
\le&c(S-S_*)(S_*I_*-S_*I+IS_*-IS)+cS_*(I-I_*)(S-S_*)\\
\le&cS_*(S-S_*)(I_*-I)+cS_*(I-I_*)(S-S_*)\\
=&0.
\ea
\]
Note that $\mv'_2=0$ implies that $S=S_*$. Any trajectory that starts in the
space $S=S_*$ and then remains in $S=S_*$ for all $t>0$ must satisfy $S'=0$, i.e.,
$I=I_*$, and similarly, we have $C=C_*, M=M_*$, and $R=R_*$. That is, the largest
positive invariant set on $\{(S, I, C, M, R)\in\mathring{\Om} : \mv'_2=0\}$ is
the singleton $\{E_*\}$. By LaSalle invariant principle, $E_*$ is globally
asymptotically stable in $\mathring{\Om}$.
\end{proof}

The last result of Theorem \ref{ThmStab2} raises the question that if the interior endemic equilibrium $E_*$ is also globally asymptotically stable for the original system \eqref{SICMR}?

\section{Application to U.S. Covid-19 Pandemic}

\bigskip\noindent
\textbf{Fit Model to Data.} The first date when the Covid-19
case and death numbers was reported
for the U.S. from CDC is Jan. 22, 2020. The end date of the data for this
study is Sept. 1, 2021 (\cite{CDC21}). To apply our model (\ref{SICMR}),
we do not expect its parameters to remain constant for this
period of the U.S. Covid-19 pandemic. We will adopt the same protocol
of best-fitting model to data from \cite{bdeng2023}. That is, starting from
day 50 (03/12/2020) to day 590 (09/1/2021), we fit the model to data from
the past 21 days. We use the
initial parameter values from \cite{bdeng2023} for $S_0, I_0,C_0,M_0$
and $c, p, a, m, d, q$, respectively, as initial guesses for the same gradient
decent algorithm as for \cite{bdeng2023} for our expanded model (\ref{SICMR}).
As for the additional parameters $\alpha$, we use a U.S. birth rate of  12.012 per 1000 per year which translates to a fixed $\alpha$ value at $\alpha=12.012/1000/365$. For parameters $\mu$, we use a U.S. death rate of  8.4 per 1000 per year which translates to a fixed $\mu$ value at $\mu=8.4/1000/365$. For parameter $\varepsilon$, we fix it at $\varepsilon=10^{-8}$. For parameter $\gamma$ we use $0.001$ as the initial guess for the best-fit searching
algorithm. On each matching day (between day 50 and day 590),
a large number of search is carried out and the best 30 results are ranked and archived. These best-fitted initials and parameters, including all the figures generated
below, are included in figshare \cite{bdeng2023b}.

\begin{figure}[t]
\vskip -.5in

\centerline
{\scalebox{.7}{\includegraphics{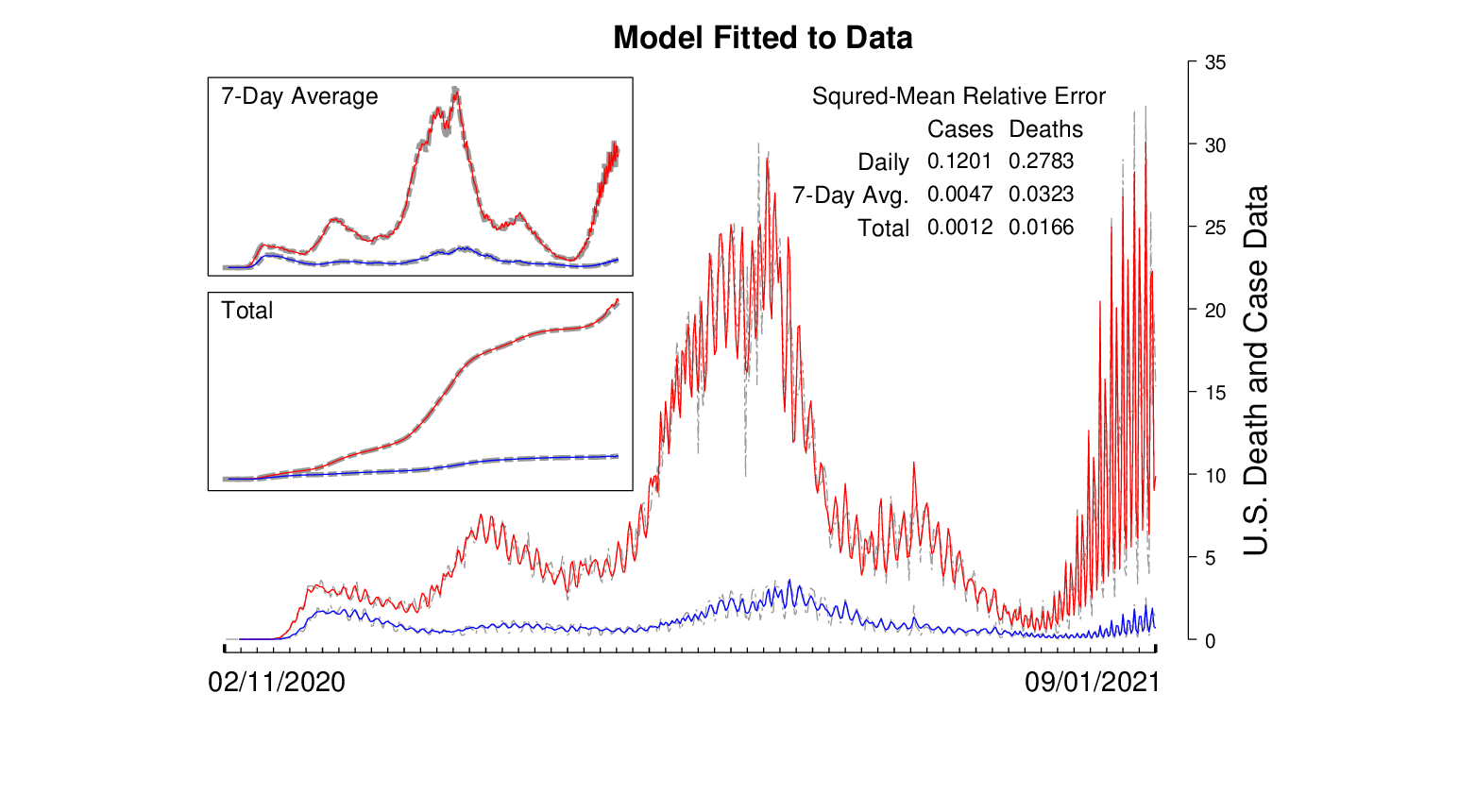}}}
\vskip -.5in
\caption{\textbf{Data-Fitting:} The unit for the
case number (data in gray and fit in red)
is 1 for $10^4$, and the unit the death number (fit in blue)
is 1 for $1.25\times 10^3$. For example, the number
4 tick-mark on the scale represents 40,000 for cases
and 5,000 for deaths. This false-scale for plot is used
to boost the visibility of the death data. All plots
use the same case-to-death plot ratio. All dot-dashed
curves (gray) are real data.} \label{figBestFit}
\end{figure}

Fig.\ref{figBestFit} shows the result of how our SICMR model is
fitted to the U.S. case and death data. The graph is assembled
by the same protocol as \cite{bdeng2023}. It uses only
the first ranked fit for each day of the 30 best-fits archived.
Specifically, for each day's case number there are
21 best-fits:
on the day the datum belongs, on the day after, up to
the 20st day after. Each day's fitting is treated
equally as every other other 20 days fitting.
Thus, each day's plotting point is
the average of the 21 best-fits. The same method is
applied to the death data and matching curve. The main graphs are
for the daily numbers, with the inserted graphs for the seven-day
average, and the cumulative total, all are computed from the daily numbers.

\begin{figure}
\vskip -.5in

\centerline
{\scalebox{.7}{\includegraphics{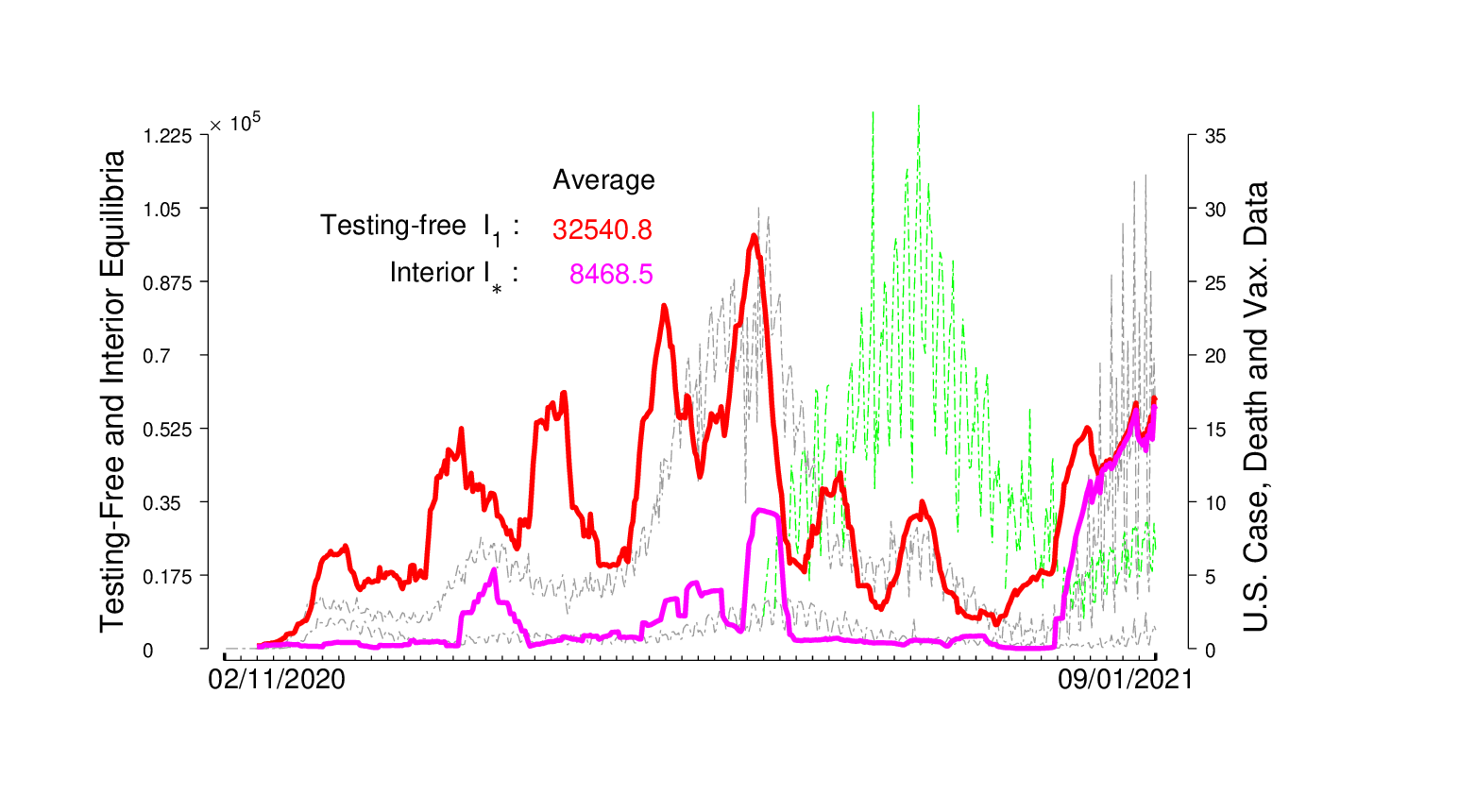}}}
\vskip -.5in
\caption{\textbf{Testing-free and Interior Equilibria:} The background
data on cases and deaths use the same scale as Fig.\ref{figBestFit}.
The scale for vaccination (green)
is 1 unit for $1.25\times 10^5$.} \label{figEquilibra}
\end{figure}

For each of the best-fit (from a total of $541\times 30$),
the best-fitted model satisfies the condition (4) of
Theorem \ref{ThmExist} which is the same as
the condition (5) of Theorem \ref{ThmStab}. Figure \ref{figEquilibra}
shows the $I$-component of the testing-free equilibrium $E_1$
and the interior equilibrium $E_\ast$. Each day's datum is
the average of 21 best-fitted values for both $I_1$ and
$I_\ast$, respectively. It shows that $I_\ast<I_1$ as predicted
by  Theorem \ref{ThmExist}(4).

\bigskip\noindent
\textbf{Variant Outbreaks.} We also know the
world was hit by the appearance of new variants of the Covid-19
virus. For this paper, we will define variants only from the
data by the underlining long term peaks of the data. For the period from day 50 to day 590, we identify 5 such peaks. The first is due to the original outbreak. The
second peaks around day 177, the third around day 347,
the fourth around day 442, and the last continues on day 590.
One can argue for only 4 variant peaks because the fourth can
be considered as a part of the third variant.

\begin{figure}[t]

\centerline
{\scalebox{.7}{\includegraphics{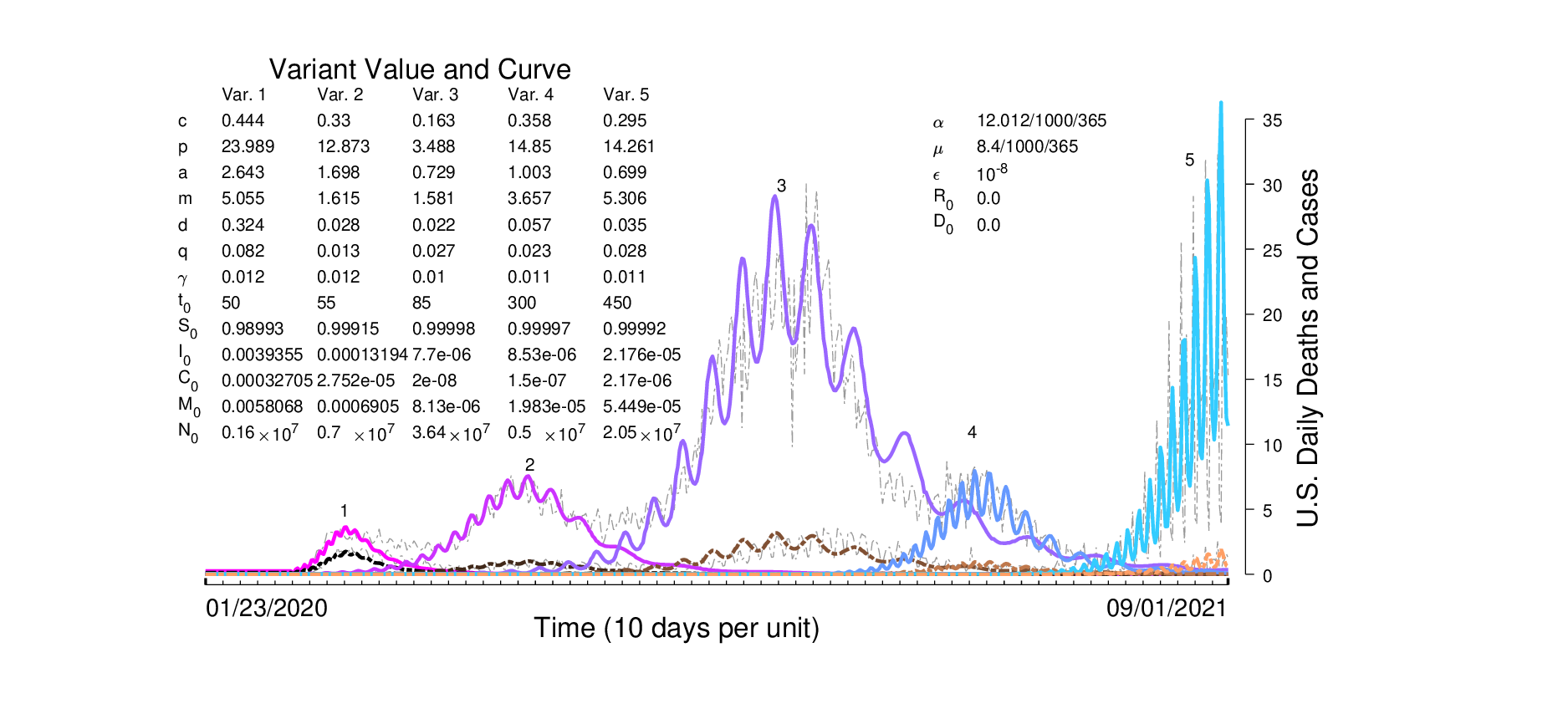}}}
\vskip -.5in
\caption{\textbf{Variants Outbreak:} The background
data on cases and deaths use the same scale as Fig.\ref{figBestFit}.
Here, $t_0$ is the initial time for a variant, $S_0$ through
$D_0$ are the initial conditions, with the new variant's
initials for the recovered class $R_0$ and the death class
$D_0$ both being zero. Parameter $N_0$ is the effective
susceptible population for a variant.}
 \label{figVariants}
\end{figure}

We used the parameter values from the best-fit of Fig.\ref{figBestFit}
to find good fits for each of the variant outbreaks.
The shared data contains 500 fits for each variant. The best-fits
are searched only for the shapes and magnitudes of the variants,
foregoing the secondary oscillation modes with 7-day and 3-day
periodicity, respectively. Figure \ref{figVariants} shows the first
ranked fit for each variant. The reason that the dimensionless
initials in $S_0$ through $M_0$ need to be accurate to the fifth
decimal place is because the daily effective susceptible population
$N_0$ is in the $10^6$ to $10^7$ range.

\begin{figure}[t]
\centerline{\parbox[b]{3.2in}
{\includegraphics[width=3.2in,height=2.4in]{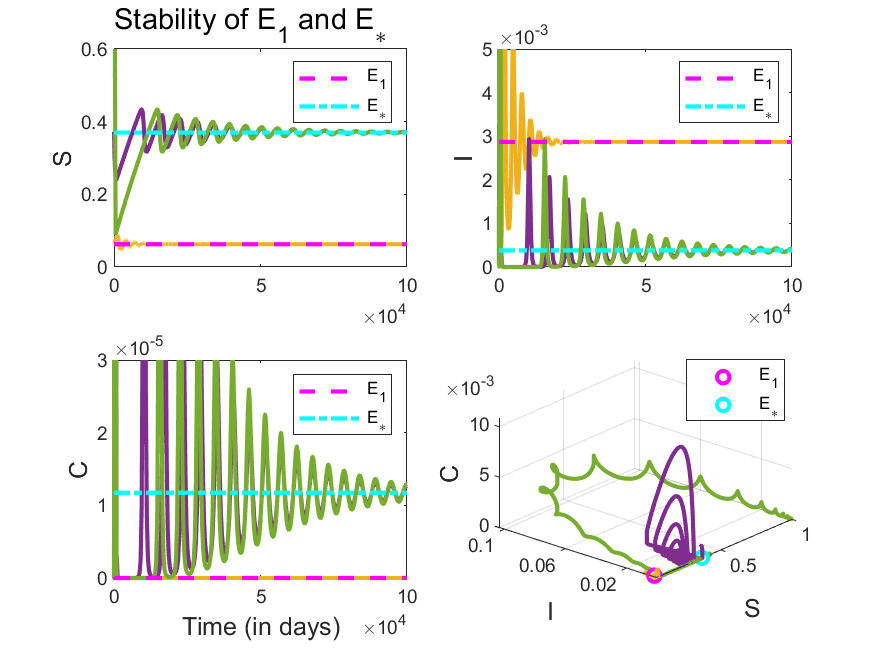}}
\parbox[b]{3.2in}
{\includegraphics[width=3.2in,height=2.4in]{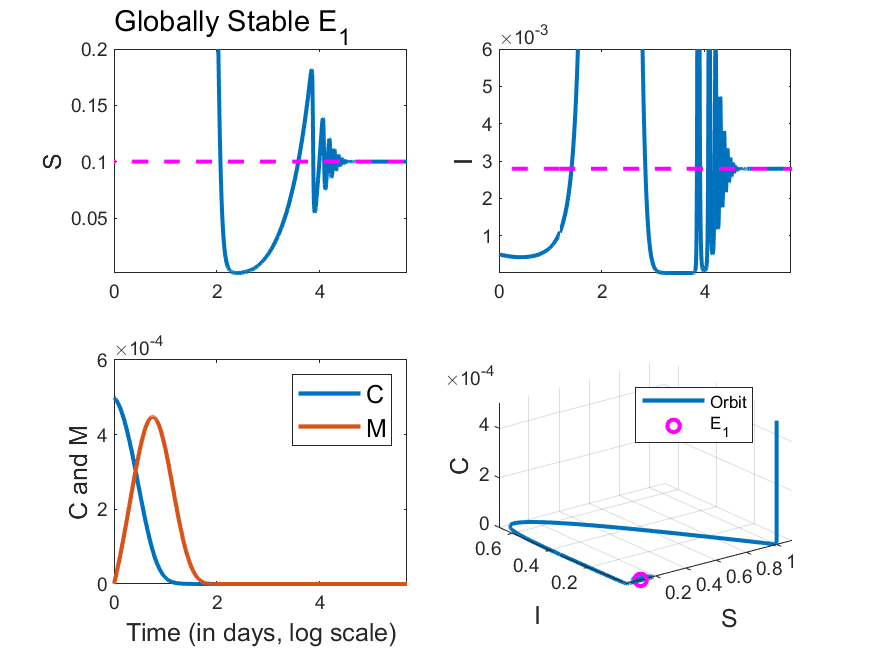}}
} \centerline{(a)\hskip 3in (b)}
\caption{\textbf{Stability of Endemic States:}
(a) Parameter values: $c=0.16,  p=3.49,  a=0.73, m=1.58, d=0.02, q=0.03,
\alpha=3\times 10^{-5}, mu=2\times 10^{-5}, \varepsilon=10^{-8}, \gamma=0.01$.
Three numerical orbits are shown: the unstable manifold orbit $W^u(E_1)$
(orange), a small perturbation orbit of $E_\ast$ (purple),
and an outbreak orbit (green). $W^u(E_1)$ converges to $E_1$ and
the outbreak converges to $E_\ast$. (b) Globally stable
$E_1$ with parameter values: $c=0.1,  p=0.4,  a=3.0,
m=0.5, d=0.01, q=0.1$ with $\alpha,\mu, \varepsilon, \gamma$ the same as (a).
The parameter values satisfy condition (4) of Theorem \ref{ThmStab}.
\label{figStabilityofE1}}
\end{figure}

\bigskip\noindent
\textbf{Local Stability of $E_1$ and $E_\ast$.} In
Fig.\ref{figStabilityofE1}(a), the parameter
values for the system \eqref{SICMR} are the same as the
variant 3 best-fit from Fig.\ref{figVariants}, rounded to two digits
in their decimals. The system satisfies the condition (4) of Theorem \ref{ThmExist}
and the condition (5) of Theorem \ref{ThmStab}.
Hence $E_1$ is unstable and there is a unique $E_\ast$. It can be demonstrated
numerically that it has at $E_1$ one negative eigenvector $-\mu$ with eigenvector $[0,0,0,0,1]$
because the $R$ equation is decoupled from the rest, two complex eigenvalues with
negative real part with eigenvectors in the invariant space $C=M=0$ for the reduced
SIR system, in which $E_1$ is globally stable. It also has one negative
eigenvalue: $-0.03$ with eigenvector
$v=(0, 0, 0, 1, -1)$, one positive eigenvalue, $1.89$, with eigenvector of
all non-vanishing entries. Denote the eigenvector by $v^u$ for the positive
eigenvalue that points into the positive side of variable $C$ and has the unit length.
As for $E_\ast$, it can be checked numerically that it is locally asymptotically stable.

Fig.\ref{figStabilityofE1}(a) shows three numerical orbits in addition
to the equilibrium solutions $E_1$ and $E_2$. The unstable manifold orbit,
denoted by $W^u(E_1)$, is generated by the initial point $X_0=E_1+10^{-5}v^u$.
The small perturbation orbit of $E_\ast$ is generated by an initial
$X_0=E_\ast+(0.01, 0.001, 0.001, 0, 0)$, and a typical outbreak orbit with
the same initial values as the variant 3
best-fit from Fig.\ref{figVariants}. (An outbreak orbit is loosely defined with
the property that the initial value of $S_0$ is near 1 while all others are
very small.) The parameter values are the same as the variant 3 fit.
The unstable manifold orbit $W^u(E_1)$ returns to $E_1$, appears to
be a homoclinic orbit.

Fig.\ref{figStabilityofE1}(b), the corresponding system \eqref{SICMR} satisfies
the condition (3) of Theorem \ref{ThmExist} and the condition (3) of Theorem \ref{ThmStab}. Hence $E_1$ globally asymptotically table and $E_\ast$ does not exist. The simulation confirms the theory.

\begin{figure}[t]
\centerline{\parbox[b]{3.2in}
{\includegraphics[width=3.2in,height=2.4in]{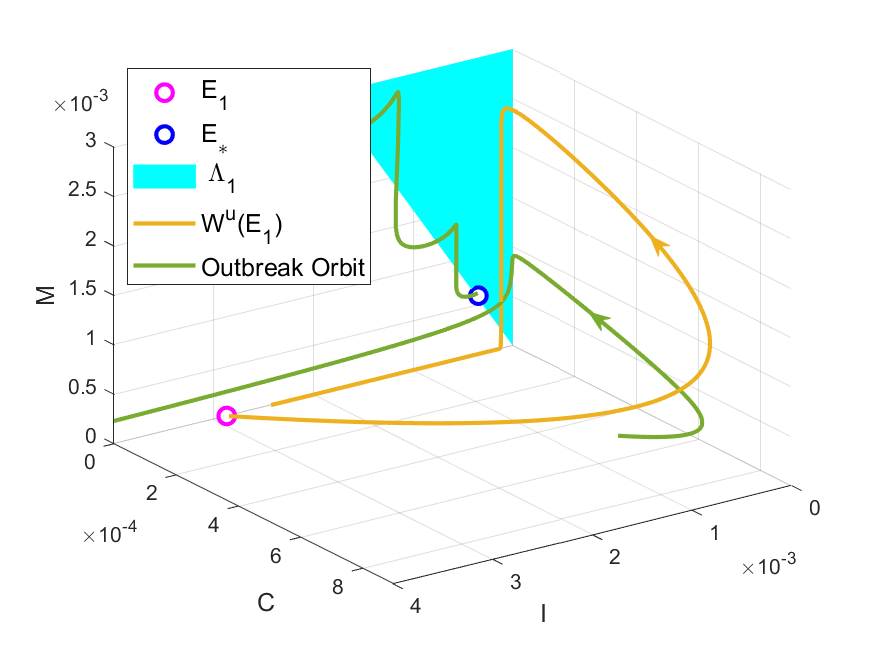}}
\parbox[b]{3.2in}
{\includegraphics[width=3.2in,height=2.4in]{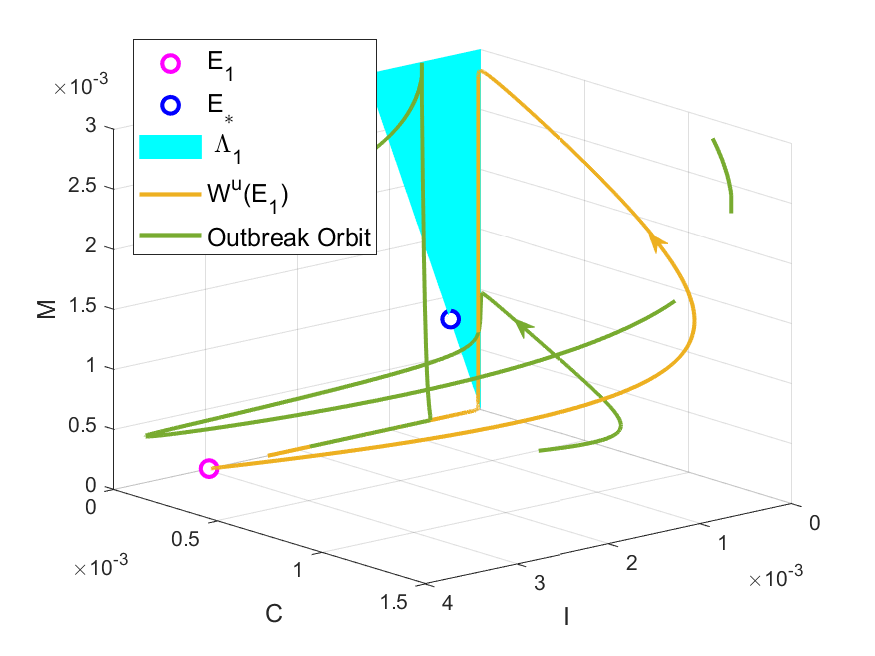}}
} \centerline{(a)\hskip 3in (b)}
\caption{\textbf{Stochastic Trapping:} (a) The invariant manifold
$\Lambda_1$ stochastically traps the unstable manifold $W^u(E_1)$
but not the outbreak orbit. (b) Stochastic trapping
for both $W^u(E_1)$ and for the variant-5's
outbreak initials from Fig.\ref{figVariants}.}
 \label{figHomoStab}
\end{figure}

\section{Stochastic Trapping and Homoclinic Connection}

Let $\Ld:=\{(S, I, C, M, R)\in\Om : C=0\}$ and $\Ld_0:=\{(S, I, C, M, R)\in\Om : C=0, M=0\}$.
Obviously, $\Ld$ is a smooth invariant manifold for the model. On it, the model is reduced to the basic SIR model with $E_1$ being globally stable with $\mr_0>1$. By Fenichel's theory of hyperbolic invariant manifolds (\cite{Feni79, Deng90, Sche08}), we can partition $\Ld$ into hyperbolic regions by finding the eigenspace at every point on $\Ld$. To do so, we first evaluate the Jacobian $J$ from the proof of Theorem \ref{ThmStab}
at $X_0=(S_0,I_0, 0, M_0, R_0)\in\Ld$
to get:
\begin{equation*}
J_0=\begin{pmatrix}-cI_0-\mu & -cS_0 & 0 & 0 &0\\
cI_0 & cS_0-\gamma-\mu & -\frac{pI_0}{\varepsilon+I_0+aM_0} & 0 & 0 \\
0 & 0 & \frac{pI_0}{\varepsilon+I_0+aM_0}-m-d-\mu & 0 & 0 \\
0 & 0 & m & -q-\mu & 0\\
0 & \gamma & 0 & q & -\mu
\end{pmatrix}.
\end{equation*}
One can check easily that it has eigenvalues: $\lambda_1=-\mu, \lambda_2=-q-\mu,
\lambda_3= \frac{pI_0}{\varepsilon+I_0+aM_0}-m-d-\mu$ and $\lambda_{4,5}$ from
the $2\times 2$ top-left block of $J_0$, which corresponds to the eigenvalues
for the reduced SIR model with $C=0$. For $\mr_0>1$, $\textrm{Re}\lambda_{4,5}<0$
because $E_1$ is asymptotically stable for the SIR system. Hence, the manifold $\Ld$
is partitioned into two open regions:
\[
\Ld_1=\Ld\cap\{\lambda_3<0\} \hbox{ and } \Ld_2=\Ld\cap\{\lambda_3>0\}.
\]
Their boundary ($\lambda_3=0$) can be solved easily to be this hyperplane:
\[
I_0=L(M_0):=\frac{(\varepsilon+aM_0)(m+d+\mu )}{p-m-d-\mu}
\]
Thus, on any interior compact subset of $\Ld_1$, the full SICMR system is
uniformly attracting, and the eigenvector, $v_3=(0, 0, 1, 0, 0)$,
for $\lambda_3$ is perpendicular to $\Ld_1$. Locally around such
a compact subset, if the eigenvalue $\lambda_3$ is greater than the others
in magnitude, then by Fenichel's theory the system admits a
hyperbolic splitting transversal to the invariant manifold, uniformly
attracting at each point, having an invariant foliation
transversal to the manifold.

Recall that the one-dimensional eigenspace of $E_1$ is
transversal to $\Ld$ with
a non-negative $C$-component. The unstable manifold
$W^u(E_1)$ is an orbit outside $\Ld$. It is called
a \textit{pseudo-homoclinic orbit} if the unstable manifold is
connected to a stable foliation of a point on $\Ld_1$
that admits a transversal hyperbolicity. Dynamics near true
homoclinic orbits can be extremely complex
(\cite{Shil70,Deng89,Chow90,Deng93,Chua01}),
we expect nontrivial dynamics near pseudo-homoclinic
orbits.

By definition, the attracting manifold $\Ld_1$ is said to \textit{stochastically trap}
an orbit outside if any numerical simulation of the orbit sinks
into the manifold with its $C$-component non-positive, $C(t)\le 0$,
for some future time $t>0$. This can happen when the orbit is
attracted to $\Ld_1$ and stays long enough near $\Ld_1$ so that the numerical
approximation of its $C$-component is indistinguishable from zero. When it happens,
a typical solver will keep $C=0$ because of the invariance of $\Ld$ to the SICMR
system. Biologically, it means that testing comes to sudden stop when
the number of confirmed $C$ is too small.

A pseudo-homoclinic orbit is called a \textit{stochastic homoclinic orbit} if the
orbit $W^u(E_1)$ is stochastically trapped by $\Ld_1$. This is what happens to
$W^u(E_1)$ for Fig.\ref{figStabilityofE1}(a) and Fig.\ref{figHomoStab}(a).
More specifically, we can see that in Fig.\ref{figHomoStab}(a) the orbit
first comes out from $E_1$, makes an U-turn, and then heads towards
$\Ld_1$. It appears to be trapped
by $\Ld_1$ because the orbit makes a right-angle
downturn following the dynamics on $\Ld$ on which $M$ is strictly decreasing with
the exponential rate $-q-\mu$, towards the sub-manifold $\Ld_0$, on which the orbit
has nowhere to go but asymptotically attracted to $E_1$ in the $SIR$ subspace.
Because of $\Ld_1$'s hyperbolicity, the trapping to the manifold is exponential
with rate $\lambda_3<0$. Thus, the farther away from the boundary of $\Ld_1$,
the greater the attraction becomes and the more likely that trapping takes place.
Stochastic trapping was confirmed empirically because all our numerical simulations had their $C$-components sink below zero even when the absolute error and relative error tolerances for the Matlab ODE solver, ode15s, were reduced all the way down to $10^{-16}$. Stochastic trapping did not happen to the outbreak orbit for higher accuracy of the solver for the outbreak initials of Fig.\ref{figHomoStab}(a).

We also carried out the same analysis for the variant-5's outbreak of
Fig.\ref{figVariants}. Stochastic trapping takes place up to $10^{-16}$
solver accuracy for both the unstable manifold of $E_1$ and the outbreak
orbit, c.f. Fig.\ref{figHomoStab}(b).

\section{Concluding Remarks}

As pointed out in \cite{Deng23},
the U.S. daily numbers exhibit a 7-day oscillation which
then changes to a 3-day oscillation. The inclusion of the Holling's Type II functional form for testing can capture this feature of the U.S. pandemic data and we failed to do
the same with the simplified model \eqref{s-sicm}. Note also
that the SICM model of \cite{Deng23} is the minimal model
to capture such oscillations at the daily scale.

Because the $E_\ast$ is globally stable for the simplified SICMR system (\ref{s-sicm})
by Theorem \ref{ThmStab2}(b), it is reasonable to conjecture the same for the
original SICMR system (\ref{SICMR}) with condition (5) of Theorem \ref{ThmStab}.
But there is a hint that may not be true. If the pseudo-homoclinic orbit
of Fig.\ref{figHomoStab} is a real homoclinic orbit,
converging to $E_1$ along the principal stable manifold tangent
to the $SI$-plane, then it is the Shilnikov's saddle-focus type,
because the real part of the stable eigenvalue is $-2.4\times 10^{-4}$,
and the unstable eigenvalue is $1.89> 2.4\times 10^{-4}$
(\cite{Shil70, Deng93, Chua01}). As a result, the dynamics
in a small neighborhood of the homoclinic orbit is chaotic, having
infinitely many periodic orbits at the minimum. For pseudo-homoclinic
orbit of the same saddle-focus type, we should expect the same. Hence the
existence of periodic orbits in a neighborhood of the orbit would
prevent the endemic equilibrium state $E_\ast$ from globally stable.
However, it remains an open problem to show the existence of chaos near
a pseudo homoclinic orbit of the Shilnikov's type.

As for the long term prospect of the U.S. pandemic, our results suggest
two possibilities. One, the outbreak is stochastically trapped to the
testing-free endemic state $E_1$, and two, the outbreak settles into
the endemic state $E_\ast$ with testing. For the latter scenario, the simulated
equilibrium in $C_\ast, M_\ast$ are approximately $1.1734\times 10^{-5},
6.1757\times 10^{-4}$, respectively, which translates to roughly
6,000 cases for $C$ and $M$ classes together each day because
the effective susceptible population $N_0$ is in the order of $10^7$.
This means, even if the endemic ends with testing, the scale is
too small to equate it with the large scale in testing we have
had throughout the pandemic. For the first
scenario, the time needed to be stochastically trapped
to the complete testing-free state $C=M=0$ is about a year, after the last
outbreak. Altogether, it suggests that testing in the U.S.
is to come to an end shortly after the end of the pandemic. This is apparent
for everyone to see but it is nonetheless surprising that the same
picture can come from a mathematical model.

\end{document}